\documentclass[11pt]{article}
\usepackage{graphicx,psfrag}
\usepackage{amsmath,amssymb}
\usepackage{epstopdf}
\usepackage{verbatim}	% to allow the use of \begin{comment}
\def\Vgz{{\gamma/Z}} 
\def\FoneVv{F_{1V}^\Vgz}
\def\FoneVa{F_{1A}^\Vgz}
\def\FtwoVv{F_{2V}^\Vgz}
\def\FtwoVa{F_{2A}^\Vgz}

\def\Fonevv{F_{1V}^v}
\def\Foneva{F_{1A}^v}
\def\Ftwovv{F_{2V}^v}
\def\Ftwova{F_{2A}^v}

 \normalsize

\newcommand{\bmat}{\left(\begin{array}}
\newcommand{\emat}{\end{array}\right)}
\newcommand{\be}{\begin{equation}}
\newcommand{\ee}{\end{equation}}
\newcommand{\bea}{\begin{eqnarray}}
\newcommand{\eea}{\end{eqnarray}}

\def\lsim{\raise0.3ex\hbox{$\;<$\kern-0.75em\raise-1.1ex\hbox{$\sim\;$}}}
\def\gsim{\raise0.3ex\hbox{$\;>$\kern-0.75em\raise-1.1ex\hbox{$\sim\;$}}}

\begin{document}
\bibliographystyle{unsrt}
\renewcommand{\thefootnote}{\fnsymbol{footnote}}
 \rightline{LAL-15-79}
\vspace{1.5cm} 
\begin{center}
{\LARGE \bf 
Probing New Physics
using top quark polarization in the $e^+e^-\to t\bar{t}$  process at future Linear Colliders\footnote{Proceeding for the series of the {\it TYL/FJPPL workshops on Top Physics at the ILC} held at KEK  (November 2013),  LPNHE Paris  (March 2014), and LAL Orsay (March 2015).} 
%Study of the polarization of the top quark with the $e^+e^-\to t\bar{t}$  process at future Linear Colliders\footnote{Proceeding for the series of the %{\it TYL/FJPPL workshops on Top Physics at the ILC} held at KEK,  LPNHE Paris, and LAL Orsay (November 2013, March 2014, March 2015).} 
}
\end{center}
\vspace{.3cm}

\begin{center} 
\sc{ \large P.H. Khiem$^{a}$\footnote{Now at Deutsches Elektronen-Synchrotron (DESY), Platanenallee 6, D-15738 Zeuthen, Germany}, E. Kou$^b$, Y. Kurihara$^a$ and F. Le Diberder$^b$} \\
\vspace{1cm}
\sl{$^a$High Energy Accelerator Research Organization (KEK),\\
 1-1 Oho, Tsukuba, Ibaraki 305-0801, Japan}\\
\sl{$^b$Laboratoire de l'Acc\'el\'erateur Lin\'eaire, IN2P3/CNRS et Universit\'e Paris-Sud 11, \\
Centre Scientifique d'Orsay, B.~P. 34, F-91898 Orsay Cedex, France }
\end{center}

\vspace{.3cm}
\vskip 0.3cm
\begin{abstract}
We investigate the 
%statistical 
sensitivity to new physics of the process $e^+e^-\rightarrow t\bar t$ 
when the top polarization is analyzed using leptonic final states $e^+e^-\rightarrow t\ \bar t\rightarrow l^+l^- b\ \bar b \ \nu_l\bar\nu_l$.
We first show that the kinematical reconstruction of the complete kinematics is experimentally tractable for this process.
Then we apply the matrix element method to study the sensitivity to the $Vt\bar t$ coupling ($V$ being a vector gauge boson), at the tree level and in the narrow width approximation.
Assuming the ILC baseline configuration, $\sqrt{S}=500\ {\rm GeV}$, and a luminosity of $500\ {\rm fb}^{-1}$, we conclude that this optimal analysis
allows to determine simultaneously the ten form factors that parameterize  the $Vt\bar{t}$ coupling, below the percent level. We also discuss the effects of the next leading order (NLO) electroweak corrections using the GRACE program with polarized beams. It is found that the NLO corrections to different beam polarization lead to significantly different patterns of contributions. 
\end{abstract}
\vskip 0.3cm  \vskip 1.2cm

%%%%%%%%%%%%%%%%%%%%%%%%%%%%%%%%%%%%%%%%%%%%%%%%%%%%%%%%%%%%%%%%%%%%%%%%%%%%%%%%%%%%%%%%%%%%%%%%%%%%%%%%%%%%
\newpage 
\section{Introduction}
\noindent
Top quark was discovered in 1995 as the 6th quark. Its heavy mass, which is orders of magnitude higher than those of the other quarks,  was predicted a decade before by indirectly measuring the radiative processes which receive a contribution from top quark propagating in the loop, which increases as the mass becomes higher. Today, a more precise measurement of the top quark mass is desired in order to discover a small deviation from the electroweak theory indirectly through loop corrections.  Furthermore, top quark mass being close to the electroweak scale, 
one can speculate that the top quark may play a special role for the electroweak symmetry breaking, namely in the new physics models. 

On the other hand, the top quark has another peculiarity: its decay time is so short that it does not hadronize.
%before decaying. 
This leads to a great advantage: the top quark is the only quark whose spin polarization can be studied.
In that sense it is akin to the tau, whose production and polarization was studied at LEP and SLC, 
but with two remarkable qualitative differences due to its large mass:
all four intermediate polarized states: $t_L\bar t_R\ , \ t_L\bar t_L\ , \ t_R\bar t_R\ ,\ t_R\bar t_L$
are produced, and interfere, therefore creating a rich pattern bearing the imprint of the top mass;
the $W$'s are produced on-shell, and their decay can be fully reconstructed.
The study of polarization allows us to probe the chirality of the interactions between the top quark and the gauge bosons $\gamma$ and $Z$  or any new particles beyond the SM.  Furthermore, the top polarization measurement also provides us an opportunity to study the CP violating interactions, in a particularly  clean manner, unhampered by hadronic effects. The use of the top polarization can be done at $e^+e^-$ colliders, such as ILC~\cite{ILC:TDR-DBD}, where massive numbers of top anti-top pairs can be produced. 
It is well known that lepton colliders, without any hadronic effect in the initial state, can provide a clean environment to study the top quark properties\cite{ILC:TDR-DBD}.  
In particular, the option of the ILC proposal to use polarized beams is quite interesting as the initial state beam polarization and final state polarization exhibit strong correlation, which can be used as an additional information to study the top quark interactions. 

The top quark polarization pattern can be reached by measuring the angular distribution of the  top quark decay products. 
For top and anti-top quarks decaying into the leptonic final states 
$e^+e^-\rightarrow t\ \bar t\rightarrow l^+l^- b\ \bar b \ \nu_l\bar\nu_l$
($l^+l^-=e^+e^-, e^+\mu^-,\mu^+ e^-,\mu^+\mu^-$) it is known that the full kinematics (hence including the top directions) can be obtained from the lepton and bottom momenta~\cite{Atwood:1991ka, Ladinsky:1992vv}. In section {\ref{Section3}} we first demonstrate that, working at tree level, one can determine the so-called form-factor of the $t\bar{t}$ production with a precision down to the few per-mill level. In this analysis, we use the {\it matrix element method} 
%(cf. appendix for a brief introduction) 
which is the most efficient method when all the kinematics can be reconstructed. 

The obtained result is quite interesting. 
Thus, it appears feasible to investigate further beyond the tree level, namely including 
the ${\mathcal{O}}(\alpha)$ and the ${\mathcal{O}}(\alpha_s)$ corrections. It has been a long time since a large  electroweak NLO correction for $e^+e^-\to t\bar{t}$ was recognized~\cite{Fujimoto:1987hu} and confirmed independently in~\cite{Fleischer:2003kk}. 
However, this effect has not yet been taken into account for the top polarization study. Since the foreseen experimental errors can be as small as per-mill level, the theoretical error coming from the higher order corrections has to be evaluated very carefully. 

The remaining of the sections is organized as follows. In section \ref{Section2}, we first revisit the feasibility of the kinematical reconstruction of the fully leptonic final states. In section \ref{Section3}, we perform a sensitivity study in terms of the effective form factors which describe the $t\bar{t}Z/\gamma$ coupling. 
Section \ref{Section4} includes the discussion on the electroweak NLO corrections, 
and the conclusions and the future prospects are given in section \ref{Section5}. 

In the following
%unless otherwise stated, 
we assume a luminosity of $500\ {\rm fb}^{-1}$ collected at a fixed center of mass energy of 500 GeV with an equal sharing in luminosity between the two beam polarizations of the ILC,  which corresponds to about $10^4$ events with ${\cal P}_{e^-}=-80\%$ and ${\cal P}_{e^+}= +30\%$, and $6\times 10^3$ events with ${\cal P}_{e^-}=+80\%$ and ${\cal P}_{e^+}= -30\%$. No acceptance cuts are applied; the reconstruction is assumed to be perfect; and detector and physics effects (ISR, hadronization, backgrounds, etc.) are ignored.

%%%%%%%%%%%%%%%%%%%%%%%%%%%%%%%%%%%%%%%%%%%%%%%%%%%%%%
%%%%%%%%%%%%%%%%%%%%%%%%%%%%%%%%%%%%%%%%%%%%%%%%%%%%%%
\section{Kinematical reconstruction of the fully leptonic final state: test with LO Monte Carlo events}
\label{Section2}
\begin{figure}[hb]
\begin{center}
\includegraphics[width=12cm, height=6cm]{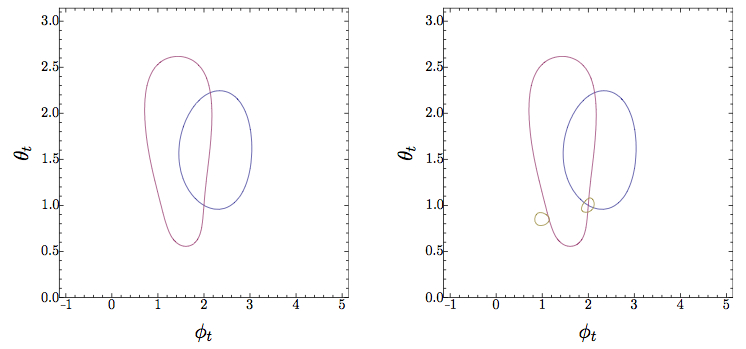}
%\includegraphics[width=6cm, height=6cm]{FirstSword.png}
%\hspace*{1cm}
%\includegraphics[width=6cm, height=6cm]{SecondSword.png}
\caption{\small The plane of the top angles in the laboratory frame; $\phi_t$ (horizontal axis), $\theta_t$ (vertical axis). The two close curves represent the solutions
of the kinematical equations for a particular event, randomly chosen. 
On the left panel, the two countours are obtained by imposing that the computed energy of  the lepton in the rest frame of the $W$ is equal to $m_W/2$.
On this panel, the ellipsoidal contour on the left (resp. right) corresponds to $l$ (resp. $\bar l$). These two countours cross in two points, which are the two solutions of the 
kinematical reconstruction.
On the right panel, 
the two small additional ellipsoidal contours are drawn by using the $b$-jet energy information:
they correspond to the domains allowed at 99\%CL.
The input values (here $\phi_t=2$ and $\theta_t=1$) are unambiguously selected by this additional constraint.}
\label{solution}
\end{center}
\end{figure}
To recover the six unknown of the three-momenta of the two missing neutrinos one can use several constraints,
in the narrow width(s) approximation: the two masses of the $t$ and $\bar t$ and the two masses of the $W^+$ and $W^-$.
With two additional constraints one can compute the six unknown needed.
In contrast to the pioneering work of ~\cite{Ladinsky:1992vv}, the treatment we propose ignores possible missing-photon(s) and
uses all four equations expressing the conservation of the four momenta.
Together with the four top and W mass constraints, 
this provides a total of eight constraints, whereas six are needed.
The two constraints in excess are used to compute also the energies of the two $b$-jets,  assumed to be poorly measured.
The directions of the two $b$-jets are assumed to be reliably determined,
even in presence of hard gluon radiation.
%\footnote{This needs to be confirmed by detailed Monte Carlo studies.}.
Solving this set of eight equations with eight unknown yields two solutions, in a first step.
These multiple solutions are further amplified when one
exchanges the role ot $b$ and $\bar b$ (assuming that the charge of the $b$-jets are not determined experimentally). 
To select the right solution, one can make use of the measurement of the $b$-jet energies. 
Although the $b$-jet energies can be only poorly measured, compared to the lepton energies,
%($\sigma (\sqrt{E_b})\simeq 0.3$~\cite{ILC:TDR-DBD})%
it turns out that one can select the correct solution in most of the cases (see Fig.~\ref{solution}).

\begin{figure}[ht]
\begin{center}
\includegraphics[width=12cm, height=6cm]{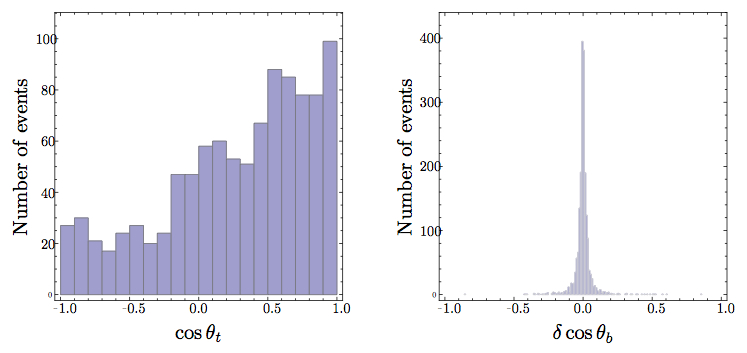}
%\hspace*{1cm}
%\includegraphics[width=6cm,  height=6cm]{DeltaCosbRec.pdf}
\caption{The left panel shows the distribution of the reconstructed $\cos\theta_t$ in the ILC rest frame obtained with the narrow width treatment
applied on Monte Carlo events produced by GRACE, at leading order, but taking into account the widths of the top and $W$.
The right panel is the distribution of the difference between the reconstructed and true values of $\cos\theta_{b}$ in the top rest frame,
here adjusting the top and $W$ masses in the final fit.
}
\label{Figcosthetatoprec}
\end{center}
\end{figure}

Using events generated by the GRACE~\cite{Yuasa:1999rg} software, a preliminary Monte Carlo study was performed
to start assessing the performance of the kinematical resconstruction for the fully leptonic case.
This (leading order) Monte Carlo simulation does not account for missing-photon(s), nor for hadronization, nor for detector effect,
but it does not rely on the narrow width approximation.

For the sake of illustration, Fig.~\ref{Figcosthetatoprec} represents the distribution of 
$\cos\theta_t$ (the reconstructed cosine of the direction of $t$ in the ILC rest frame).
The same figure (right hand side) shows the distribution of the difference between the reconstructed and true values of $\cos\theta_{b}$ (the cosine of the direction of the b with respect to the top direction, in the top rest frame) 
using Breit-Wigner constraints to let the masses of top and $W$ vary in the final stage of the fit,
which improves the precision of the reconstruction.

For about 5\% of the events, the wrong $b$ combination is retained. A significant fraction of the latter are irreducible errors: the wrong combination
being indeed a better match, by accident, due to top's and/or $W$'s being far off-shell.

%%%%%%%%%%%%%%%%%%%%%%%%%%%%%%%%%%%%%%%%%%%%%%%%%%%%%%
%%%%%%%%%%%%%%%%%%%%%%%%%%%%%%%%%%%%%%%%%%%%%%%%%%%%%%
\section{Expected sensitivity to the form factors
% using the matrix element method
: LO example} 
\label{Section3}
\subsection{Narrow Width Approximation}
In the narrow width approximation, 
and at born level, 
the distribution of events can be described in the helicity formalism.
We denote $\lambda_x$ the helicity of particle $x$.
\begin{itemize}
\item{}
The matrix element of the production of the $t\bar t$ pair through $Z/\gamma$ is denoted as : ${\cal M}_{\lambda_{e}\lambda_{\bar e}}^{\lambda_t,\lambda_{\bar t}}$,
\item{}
The matrix element of the decay $t\rightarrow l^+\nu_l\ b$ : ${\cal M}^{\lambda_t}_{\lambda_{\bar l},\lambda_\nu,\lambda_b}$,
\item{}
The matrix element of the decay $\bar t\rightarrow l^-\bar \nu_l\ \bar b$ : ${{\cal M}}^{\lambda_{\bar t}}_{\lambda_l,\lambda_{\bar \nu},\lambda_{\bar b}}$.
\end{itemize}

For fully polarized beams, the distribution of events is described by:
\begin{equation}
\mid{\cal M}_{\lambda_{e}\lambda_{\bar e}}\mid^2 = 
\sum_{\lambda_b,\lambda_{\bar b},\lambda_{\bar l}\lambda_l,\lambda_\nu,\lambda_{\bar\nu}}
\mid
\sum_{\lambda_t, \lambda_{\bar t}}
{\cal M}_{\lambda_e,\lambda_{\bar e}}^{\lambda_t,\lambda_{\bar t}}
{\cal M}^{\lambda_t}_{\lambda_{\bar l},\lambda_\nu,\lambda_b}
{{\cal M}}^{\lambda_{\bar t}}_{\lambda_l,\lambda_{\bar \nu},\lambda_{\bar b}}
\mid^2
\end{equation}

In the following we assume
that the data analysis aims to probe the couplings of the top to the $Z/\gamma$.
The coupling of the $W$ is assumed to be the one of the Standard Model (SM):
for the massless $b$, $l$ and $\nu$ only the left-handed helicities are involved, and the first sum can be removed.

%%%%%%%%%%%%%%%%%%%%%%%%%%%%%%%%%%%%%%%%%%%%%%%%%%%%%%
\subsection{Form factors and the angular distribution}
By using the angular distribution of the top quark production and decays, which can be obtained by measuring the kinematics of the fully leptonic decays, as discussed in the previous section, we can extract the top quark polarization pattern. In particular, this pattern carries the information of the interactions which produce the top and anti-top pair. In~\cite{Kane:1991bg}, assuming that the $t\bar{t}$ pair is produced from annihilation of $e^+e^-$ into a spin one particle, the most general Lagrangian for the top and anti-top production are obtained in terms of the form factors as: 
 \be
 \label{LagrangianLevel}
 \mathcal{L}_{\rm int}=\sum_{v=\gamma,Z}g^v\left[ 
 V^v_l \bar{t}\gamma^{l}(\Fonevv+\Foneva\gamma_5)t
 +\frac{i}{2m_t}\partial_\nu V_l \bar{t}\sigma^{l\nu}(\Ftwovv+\Ftwova\gamma_5) t
 \right] 
 \ee
 Since for the EDM term, the coupling $\FtwoVa$ can be a complex number, we have ten real form factors.
 % in total. 
 Note that in the literature, one can find different notation using the form factors $\tilde{F}$'s  (see e.g.~\cite{Amjad:2013tlv}). Our numerical result is partially given in terms of $\tilde{F}$'s for comparison but  these two definitions are related via simple formulae: 
 \[\tilde{\Fonevv}=-(\Fonevv+\Ftwovv), \quad \tilde{\Ftwovv}=\Ftwovv, \quad \tilde{\Foneva}=-\Foneva, \quad \tilde{\Ftwova}=-i\Ftwova \]
 By using these form factors, the angular distribution for each combination of the initial and the final polarization stems from the sum of the $\gamma$ and $Z$ exchange amplitudes (see e.g.~\cite{Schmidt:1995mr}): 
\bea
\label{lrLR}
{\mathcal{M}}(e_L\bar{e}_R\to t_L\bar{t}_R)^\Vgz &=& c_L^\Vgz [\FoneVv-\beta \FoneVa +\FtwoVv](1+\cos\theta)e^{-i\phi} \\
{\mathcal{M}}(e_L\bar{e}_R\to t_R\bar{t}_L)^\Vgz &=& c_L^\Vgz [\FoneVv+\beta \FoneVa+\FtwoVv](1-\cos\theta)e^{-i\phi} \\
{\mathcal{M}}(e_L\bar{e}_R\to t_L\bar{t}_L)^\Vgz &=& c_L^\Vgz \gamma^{-1}[\FoneVv+\gamma^2(\FtwoVv+\beta \FtwoVa)]\sin\theta e^{-i\phi} \\
{\mathcal{M}}(e_L\bar{e}_R\to t_R\bar{t}_R)^\Vgz &=& c_L^\Vgz \gamma^{-1}[\FoneVv+\gamma^2(\FtwoVv-\beta \FtwoVa)]\sin\theta e^{-i\phi} \\
{\mathcal{M}}(e_R\bar{e}_L\to t_L\bar{t}_R)^\Vgz &=& -c_R^\Vgz [\FoneVv-\beta \FoneVa+\FtwoVv](1-\cos\theta)e^{i\phi} \\
{\mathcal{M}}(e_R\bar{e}_L\to t_R\bar{t}_L)^\Vgz &=& -c_R^\Vgz [\FoneVv+\beta \FoneVa+\FtwoVv](1+\cos\theta)e^{i\phi} \\
{\mathcal{M}}(e_R\bar{e}_L\to t_L\bar{t}_L)^\Vgz &=& c_R^\Vgz \gamma^{-1}[\FoneVv+\gamma^2(\FtwoVv+\beta \FtwoVa)]\sin\theta e^{i\phi} \\
\label{rlRR}
{\mathcal{M}}(e_R\bar{e}_L\to t_R\bar{t}_R)^\Vgz &=& c_R^\Vgz \gamma^{-1}[\FoneVv+\gamma^2(\FtwoVv-\beta \FtwoVa)]\sin\theta e^{i\phi} 
\eea
where $\beta^2=1-4m_t^2/s, \gamma=\sqrt{s}/(2m_t)$ and the overall factors $c_{L/R}^{\gamma/Z}$ are: 
\be
c_L^\gamma=-1, \quad c_R^\gamma=-1, \quad c_L^Z=\left(\frac{-1/2+s_w^2}{s_wc_s}\right)\left(\frac{s}{s-m_Z^2}\right), \quad c_R^Z=\left(\frac{s_w^2}{s_wc_s}\right)\left(\frac{s}{s-m_Z^2}\right)
\ee
where $s_w=\sin\theta_w$ and $c_w=\cos\theta_W$, with $\theta_W$ being the weak mixing angle. 

The core of the experimental technique advocated here 
is to use the complete kinematics of the final state to perform a likelihood analysis 
based on the complete PDF as provided by the matrix element of the process.
Doing so, the analysis is optimal, since it uses all available information.

%%%%%%%%%%%%%%%%%%%%%%%%%%%%%%%%%%%%%%%%%%%%%%%%%%%%%%
\subsection{Optimal data analysis}
Let's define $\mid {\cal M}\mid^2(\alpha)$ the complete matrix element squared, for a given beam polarizations.
It depends on $k$ constants, here collectively denoted by $\alpha$.
Upon integration over phase space, and for a given luminosity ${\cal L}$, one expects a number of events:
\begin{equation}
\label{Normalization}
N(\alpha)={\cal L}\int \mid {\cal M}\mid^2(\alpha)\ {\rm d}{\rm Lips}
\end{equation}
where the Lorentz invariant phase space is:
\begin{equation}
{\rm d}{\rm Lips}\propto {\rm d}\cos\theta_t\ {\rm d}\cos\theta_b\ {\rm d}\phi_b\ {\rm d}\cos\theta_{\bar b}\ {\rm d}\phi_{\bar b}\ {\rm d}\cos\theta_{l^+}\ {\rm d}\phi_{l^+}\ {\rm d}\cos\theta_{l^-}\ {\rm d}\phi_{l^-}\ {\rm d}q^2_t\ {\rm d}q^2_{\bar t}\ {\rm d}q^2_W\ {\rm d}q^2_{\bar W}
\end{equation}
where the angles are defined in the appropriate rest frames, and the $q^2$'s are the invariant masses of the top's and W's.
Note that, 
since the normal to the plane defined by the $e^+e^-$ pair and the $t\bar t$ pair can be used as a reference 
to measure the azimuthal angles, all angles involved in the 6 particle final state a priori contribute to $\mid{\cal M}_{\lambda_{e}\lambda_{\bar e}}\mid^2 $,
up to an overall rotation around the $z$ axis that allows to set $\phi_t=0$, which therefore does not appear in ${\rm d Lips}$.
%\footnote{Beam polarizations are assumed to be longitudinal, not transverse: in the latter case, $\phi_t$ would remain.}.
In the following illustration, one uses the narrow width approximation, so that the four $q^2$ are integrated out. 
As a result the Lorentz invariant phase space is reduced to nine-dimension.

If one is using the full matrix element squared in a likelihood analysis relying only on the distributions of events in phase space,
the inverse of the expected covariance matrix can be expressed as:
\begin{eqnarray}
\label{MasterFormula}
V_{ij}^{(-1)}
\equiv \Lambda_{ij}
&=&
N\ 
\langle (\omega_i-\Omega_i)(\omega_j-\Omega_j)\rangle_{_0}
\end{eqnarray}
where:
\begin{eqnarray}
N&=&N(\alpha_0)
\\
\omega_i&=&{\partial\mid {\cal M}\mid^2(\alpha)\over\partial \alpha_i}_{\mid\alpha^0}{1\over \mid {\cal M}\mid^2(\alpha^0)}
\\
\Omega_i&=&{\partial N(\alpha)\over\partial \alpha_i}_{\mid\alpha^0}{1\over N(\alpha^0)}=\int \omega_i \ f_0\ {\rm dLips}
=\langle \omega_i\rangle_{_0}
\\
f_0&=&{\mid {\cal M}\mid^2(\alpha^0)\over\int \mid {\cal M}\mid^2(\alpha^0)\ {\rm dLips}}
\end{eqnarray}
where $\alpha_0$ stands for the set of $\alpha_i$ values that maximize the likelihood.
In the following we assume that they correspond to the null-hypothesis: $\alpha^0=0$ in the SM. 
To take into account the constraints coming from the yields
amounts to ignore the $\Omega_i$ and use, instead of Eq.(\ref{MasterFormula}):
\begin{eqnarray}
\label{MasterFormulaYields}
V_{ij}^{(-1)}\equiv
\Lambda_{ij}&=&
N\ \langle \omega_i\omega_j\rangle_{_0}
\end{eqnarray}
The above expressions are illustrated in appendix.
In the following, the analysis is assumed to make use of the yields (i.e. Eq.(\ref{MasterFormulaYields}) is used) which contributions are important.
For the sake of clarity, only the yields of the fully leptonic final state are used, and not the yields from other final states.

The above Eqs.(\ref{MasterFormula})-(\ref{MasterFormulaYields}) convey an important set of messages:
\begin{itemize}
\item For each theoretical parameter $\alpha_i$, and for a particular experimental set-up
(e.g. for various beam polarizations or energies) there is a corresponding 
%conjugate 
kinematical variable $\omega_i$ that captures all the relevant information carried by each event~\cite{Atwood:1991ka,Davier:1993}: therefore, this 
optimal variable should be accurately measured,
%\footnote{
%These $\omega_i$ variables were introduced in~\cite{Atwood:1991ka,Davier:1993}.
%Zerwas:1995,Bernreuther:1996a,Bernreuther:1996b,Grzadkowski:1997,Brzezinski:1999,Atwood:2000}
%},
\item The 
%conjugate 
$\omega_i$
variables can be defined in any case: the matrix element does not have to depend linearly on the theoretical parameters,
nor does it need to be cast into a readable expression (the LO expression is already quite intricate),
\item Different experimental set-up correspond to different 
$\omega_i$ variables, however, to combine the data set is straightforward: one just needs to add the corresponding
$\Lambda$ matrix to obtain the full $\Lambda$ matrix,
\end{itemize}

It should be pointed out that the use of the 
%conjugate 
$\omega_i$ variables is done implicitly when one performs a likelihood analysis using the full matrix element squared.
These 
%conjugate 
variables do not bring additional statistical power, but they provide a means to grasp more easily what the fit is doing, and which events are the important ones.

%%%%%%%%%%%%%%%%%%%%%%%%%%%%%%%%%%%%%%%%%%%%%%%%%%%%%%
\subsection{Result using the optimal analysis method}
As a matter of illustration, if one performs a simultaneous fit of the deviations with respect to the Standard Model values of 
10 of the form factors
%\footnote{For the sake of comparison we use the $\tilde F$ notations for the form factors, as used in~\cite{Amjad:2013tlv}.
%Although they are simply related to the $F$ form factors, the correlations in the fit being different, one cannot compare directly the results using %different notations.} 
one gets the following error matrix
(diagonal terms are the statistical uncertainties, while off-diagonal terms are the correlation coefficients):

\setcounter{MaxMatrixCols}{20}
\begin{minipage}{\linewidth}\small
\begin{equation}
\nonumber
% \begin{array}{*10{c}}
 \begin{bmatrix}
 {\cal R}{\rm e}\ \delta\tilde F_{1V}^\gamma & 
 {\cal R}{\rm e}\ \delta\tilde F_{1V}^Z &
 {\cal R}{\rm e}\ \delta\tilde F_{1A}^\gamma &
 {\cal R}{\rm e}\ \delta\tilde F_{1A}^Z &
 {\cal R}{\rm e}\ \delta\tilde F_{2V}^\gamma &
 {\cal R}{\rm e}\ \delta\tilde F_{2V}^Z &
 {\cal R}{\rm e}\ \delta\tilde F_{2A}^\gamma &
 {\cal R}{\rm e}\ \delta\tilde F_{2A}^Z &
 {\cal I}{\rm m}\ \delta\tilde F_{2A}^\gamma &
 {\cal I}{\rm m}\ \delta\tilde F_{2A}^Z \\
     0.0037 &  -0.18    & -0.09  & +0.14   &+0.62  &-0.15    & 0  & 0  & 0 & 0  \\
               &    0.0063  & +.14   & -0.06    &-0.13   &+0.61   & 0   & 0 & 0 & 0  \\ 
               &               & 0.0053  & -0.15   &-0.05  &+0.09    & 0   & 0 & 0 & 0  \\ 
               &               &            &   0.0083  &+0.06   &-0.04 & 0   & 0 & 0 & 0  \\          
               &               &            &              &0.0105  &-0.19    & 0   & 0 & 0 & 0 \\   
               &               &            &              &           &0.0169   & 0   & 0 & 0 & 0 \\      
               &               &            &              &           &            & 0.0068  &-0.15 & 0 & 0  \\       
               &               &            &              &           &            &            &0.0118 & 0  & 0  \\          
               &               &            &              &           &            &            &          & 0.0069 & -0.17  \\      
               &               &            &              &           &            &            &          &           & 0.0100  \\      
\end{bmatrix}             
%\end{array}
\end{equation}
\end{minipage}  

\noindent
The correlation coefficients between the 6 first form factors and the last 4 form factors exactly
vanish: this absence of correlation,
comes from the property of the integral,
when $\omega_i$ is CP even and $\omega_j$ is CP odd (see also the appendix):
\begin{equation}
\Lambda_{ij}=\langle \omega_i\omega_j\rangle_{_0}=\int \omega_i\omega_j f_0\ {\rm dLips}=0
\end{equation}
because $\int f_0\ {\rm dLips}$ is CP even. 
Similarly, the correlation between the real parts and the imaginary parts of $\delta\tilde F_{2A}^\gamma$ and $\delta\tilde F_{2A}^Z$ 
vanish because of $C$ (and $P$) symmetry.

It should be stressed that the above matrix corresponds to the ideal case: 
 detailed studies require to take into account the dilution of information due to detector and physics effects, as well as related systematical errors.
 Note that, while the beam polarizations help to improve the measurements, they are not essential for the method to apply.

The semi-leptonic decays, in principle, can also be analyzed using the matrix element method.
In that case, one trades a well measured charged lepton, but a missing neutrino with two jets, not so well measured and for which charge information is lost
(in general), and one also gains a factor 6 in statistics.
%The kinematics can also be reconstructed from the 8 constraints, using only the three-momentum of the lepton and the directions of the 4 jets to %recover
%the energies of the 4 jets, and the three momentum of the missing neutrino, and the energy of an ISR photon.
Assuming that the $b-\bar b$ jets can be properly assigned to the top decays, one can compute the matrix element squared, symmetrized with respect to
the two quarks from the $W$ hadronic decay. 
The resulting statistical errors are about a factor two smaller than for the fully leptonic final state, here also for a perfect detector.

%%%%%%%%%%%%%%%%%%%%%%%%%%%%%%%%%%%%%%%%%%%%
%%%%%%%%%%%%%%%%%%%%%%%%%%%%%%%%%%%%%%%%%%%%
\section{NLO matrix element study}
\label{Section4}
\begin{figure}[ht]
\hspace*{-.0 cm}
\centerline{
\includegraphics[width=8cm]{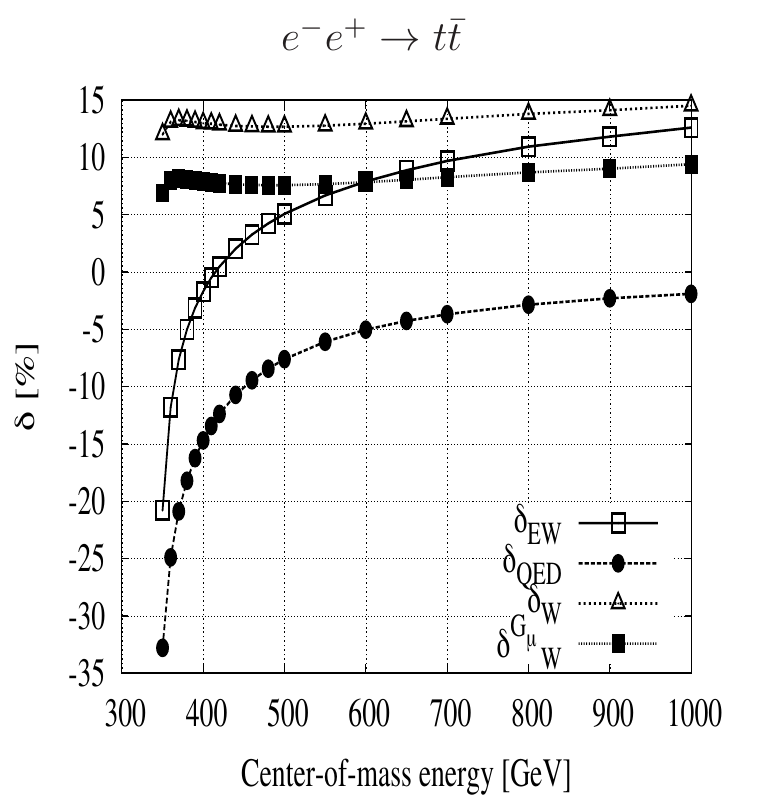}}
\caption{\small The result of 
the full electroweak correction and the genuine weak correction on the total cross section,
as a function 
of the center-of-mass energy (taken from \cite{Khiem:2012bp}).
The circle points represent the QED correction, the empty rectangle points are the results for the full electroweak correction while the triangle points are the results for the genuine weak correction in the $\alpha$ scheme. The filled rectangle points represent the results of the genuine weak correction in the $G_\mu$ scheme} 
\label{GRACE}
\end{figure}

As mentioned in the introduction, the NLO corrections to the $e^+e^-\to t\bar{t}$ process are large: 
they amount to $\sim 5 \%$ for the total cross section and $\sim 10 \%$ for the forward-backward asymmetry. 
Since the experimental measurements 
can reach the per-mill level, it is clear that we can actually measure these electroweak corrections and furthermore identify the deviation from the SM at this order. 
The GRACE program~\cite{Yuasa:1999rg} can provide the SM prediction for $e^+e^-\to t\bar{t}$ including the full one loop electroweak corrections. About 150 diagrams have been  computed  automatically~\cite{Khiem:2012bp}. 
As an example, we present the result for the total cross section of the $e^+e^-\to t\bar{t}$ process in Fig.~\ref{GRACE}. From the figure, we can extract the following conclusions:
among large corrections of the full $O(\alpha)$ terms, the largest correction comes from initial state photon-radiations, which is denoted as $\delta_{QED}$ in the figure.  
Besides the initial-state photonic-correction and another $5\%$ of trivial  weak corrections estimated using  the $G_\mu$ scheme, still $5\%$ corrections from non-trivial weak-correction remain. 

\begin{figure}[h]
\begin{center}
\psfrag{costheta}{$\cos\theta$}
\includegraphics[width=14cm]{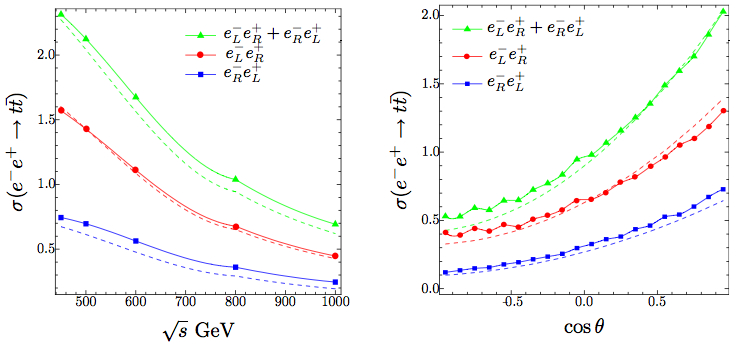}
\caption{\small Preliminary results produced by the GRACE software with polarized beams. The red (circle) and blue (square) lines are the result for initial state polarization $e^-_Le^+_R$ and $e^-_Re^+_L$, respectively. The solid lines represent the tree level plus one loop corrections while the dashed lines are tree level only. The green (triangle) lines are the sum of the red (circle) and blue (square) lines. The left panel is the $\sqrt{s}$ dependence of the cross section and the right panel is the $\cos\theta$ dependence. Note that the drawn lines are only to guide your eyes. }  
\label{GRACEII}
\end{center}
\end{figure}

Recently the initial and the final state polarizations have been implemented into the GRACE program, 
which allows us to perform a more detailed study in this direction. We show two examples of the results  obtained by this GRACE version in Fig.~\ref{GRACEII}, where we investigate the NLO contributions for the polarized initial states, $e^-_Le^+_R$ and  $e^-_Re^+_L$.  The left panel shows the NLO contribution to the cross section (solid lines) for $e^-_Le^+_R$ (red circle) and  $e^-_Re^+_L$ (blue square). Dashed lines are the tree level result and the green lines (triangle) are the sum of two polarization cases. The NLO correction is significantly larger for the $e^-_Re^+_L$ polarization case. The right panel shows the $\cos\theta$  dependence of the NLO contributions (with the same color scheme as the left panel).  Interestingly, we find that in the case of  $e^-_Le^+_R$, the NLO contributions are negative (positive) for positive (negative) $\cos\theta$ while for $e^-_Re^+_L$, the NLO contributions are always positive. 
This kind of strong dependence of the NLO terms on the kinematical variables can be most useful to investigate the NLO corrections in detail. 

It is important to emphasize that the angular distribution formalism used in the previous section does not apply fully to the NLO electroweak corrections. 
For example, in Eqs.(\ref{lrLR})-(\ref{rlRR}) it is assumed that $e^-e^+$ annihilate into spin one particles, but the box diagram which appears as a one loop correction does not fit into this category. 
 As a result, the angular distribution corrections from the box diagram cannot be written in terms of $(1\pm \cos\theta)$ or $\sin\theta$ as in the  above expressions.  
Therefore it is necessary to define a set of 
theoretical parameters that describes completely the NLO matrix element and use this full NLO matrix element in the analysis. 
Then, the numerical results obtained in the previous section are likely be affected. 
Thus, the analysis including the full NLO corrections is 
essential to be conclusive to the sensitivity to the new physics in this process. 

 %%%%%%%%%%%%%%%%%%%%%%%%%%%%%%%%%%%%%%%%%%%%%%%%%%%%
 %%%%%%%%%%%%%%%%%%%%%%%%%%%%%%%%%%%%%%%%%%%%%%%%%%%%
 \section{Conclusion and future prospects}
 \label{Section5}
In this paper, we showed that fully leptonic reconstruction of $t\bar{t}$ events 
($e^+e^-\rightarrow t\ \bar t\rightarrow l^+l^- b\ \bar b \ \nu_l\bar\nu_l$)
at linear colliders
presents a strong potential to perform precise measurements of the $t\bar{t}V$ coupling, 
complementary to the measurements provided by the semi-leptonic process.
In particular, the full kinematical reconstruction in this process allows us to successfully apply the matrix element method. 
We have performed a sensitivity study to ten form factors which parameterize in rather general terms the $t\bar{t}V$ couplings at the tree level.    
Based on the ILC baseline configuration, $\sqrt{s}=500$ GeV, with a luminosity of $500$ fb$^{-1}$,
%and with electron and positron polarization, ${\cal P}_{e^-}=\pm 80\%$ and ${\cal P}_{e^+}= \mp 30\%$, 
we found that the ten form factors can be simultaneously determined below a percent level precision. 
The most important next step is to include experimental and physics effects
to obtain a more realistic assessment of the statistical power of this analysis,
and to evaluate systematical uncertainties, both experimental and theoretical. 
For instance, one should perform the analysis using the leading order GRACE Monte Carlo producing 6-fermion final state,
not only through $t\bar t$ pair production.

We have also discussed the NLO electroweak corrections to $e^+e^-\to  t \bar{t}$,
known to be non-negligible in this process. After removing the reducible corrections (i.e. QED photon emission, corrections to Fermi constant), we find that there is still $~5$\% of loop effect remaining.  The high sensitivity demonstrated at LO by the study presented in this paper provides a proof of principle that the method could be applied, and encourages us to attempt an electroweak precision test using this process with full NLO electroweak corrections included. We have shown a preliminary result of the NLO computation of GRACE with polarized beams. Interestingly, we have found that the pattern of the NLO corrections are very different for the different configurations of the beam polarizations.
 These correlations must be important for the precise determination of the top electroweak couplings.  
 
Our final goal is to perform a similar analysis including fully the NLO corrections. 
However, a calculation for full one-loop corrections with six-body final states, including three-body decays of each top quark, 
is presently beyond our reach.
The number of diagrams involved in this process is larger than 90,000 and some of them may imply more than 100GB of executable modules. 
On the other hand, the on-shell approximation for the NLO calculation is expected to ensure enough precision for the analysis. Full one-loop electroweak corrections are estimated separately for the top-quark pair-production  and for their three-body decays, then they are convoluted as
${\cal M}_{\rm 6-body}={\cal M}_{tt}^{\rm NLO}\otimes {\cal M}_{\rm decay}^{\rm Tree}+{\cal M}_{tt}^{\rm Tree}\otimes {\cal M}_{\rm decay}^{\rm NLO}$ . 
Non-factorizable corrections, 
such as a photon bridge between initial electron and final lepton, can be estimated separately, and accounted for, if necessary.
It is important to emphasize that the form factor parameterization, thus the angular distributions, which we have used at the LO study would be inapplicable for the NLO study due to the new type of diagrams which can not be parameterized solely by the effective $t\bar{t}V$ coupling, such as the box diagrams. It would be very interesting to see how much the numerical results obtained in this article would be affected by these contributions.

\section*{Acknowledgements}
  The present study was developed in the framework of the TYL France-Japan "virtual laboratory" (LIA),
which essential support is warmly acknowledged. The authors thanks all the members of the ILC-Top team of TYL for the very lively atmosphere of the workshops. In particular, this study benefited from enlightening comments from K. Fujii, R. P\"oschl and F. Richard. 
We are greatful to J. A.M Vermaseren,  Mr. Otto Rottier (NIKHEF) and T. Ueda (KIT) for useful discussions and comments.
We would like to thank P. Janot for informing us of his recent article~\cite{Janot:2015yza}, which deals with a related subject,
and we thank J. Hebinger for his careful reading of this document.

\section*{Appendix: Illustration of the The Matrix Element Method : CP violation}
%%%%%%%%%%%%%%%%%%%%%%%%%%%%%%%%%%%%%%%%%%%%%%%%%%%%%%
\label{CPviolation}

We consider here the special case where only a subset of CP-violating parameters are intended to be measured,
and we derive step by step the result of Eq.(\ref{MasterFormula}):
%We consider the CP-violating complex parameter 
\begin{equation}
\alpha=\alpha_r+i \alpha_i=-2i\ F_{2A}^Z\sin\theta_W
\end{equation}
The matrix element can be written as the quadratic expression:
%(omitting the index $Z$ for the sake of brevity):
\begin{equation}
\mid{\cal M}_{\lambda_{e}\lambda_{\bar e}}\mid^2=O_0+O_r \alpha_r+O_i \alpha_i+O_2 (\alpha_r^2+\alpha_i^2)
\end{equation}
where the cross term $\alpha_r \alpha_i$ vanishes.
The above expression being positive definite for any values of $\alpha_r$ and $\alpha_i$, the coefficients satisfy:
\begin{eqnarray}
O_o&>&0
\\
O_2&>&0
\\
4 O_o O_2&>& O_r^2+O_i^2
\end{eqnarray} 
In particular $\mid O_{r/i}\mid>O_o$ is allowed (and does occur).
Having in mind that $\alpha_{r/i}\ll 1$ one may be tempted to ignore the $\alpha_{r/i}^2$ contributions.
However doing so does not bring any simplification, and it raises the concern of events having $O_2\gg \mid O_{r/i}\mid$:
hence, the $\alpha_{r/i}^2$ terms are kept in the following.
This illustrates the fact that the use of 
%conjugate 
the $\omega_i$ variables is not restricted to a linear dependence of the matrix element on the theoretical parameters to be measured.

Upon integration over the phase space, one gets:
\begin{equation}
\int \mid{\cal M}_{\lambda_{e}\lambda_{\bar e}}\mid^2{\rm dLips}={\cal O}_o+{\cal O}_r \alpha_r+{\cal O}_i \alpha_i+{\cal O}_2 (\alpha_r^2+\alpha_i^2)
\end{equation}
where the second and third term vanish: 
\begin{equation}
{\cal O}_r = {\cal O}_i=0
\end{equation}
This is because $O_r$ and $O_i$ are both odd under CP: then upon integration over the phase space, each point will match it CP conjugate point to cancel out the resulting integral: stated differently, in the case at hand the derivatives with respect to the theoretical parameters $\alpha$ of the yield $N(\alpha)$ vanish, for $\alpha_r^0=\alpha_i^0=0$.

The log-likelihood expression 
reads:
\begin{eqnarray}
\label{chi2NotExtended}
{\rm L}(\alpha_i)
&=&
N
\int\ln
\left(
{O_0+O_r \alpha_r+O_i \alpha_i+O_2 (\alpha_r^2+\alpha_i^2)\over {\cal O}_o+{\cal O}_2 (\alpha_r^2+\alpha_i^2)}
\right)
{O_o\over{\cal O}_o}
\ {\rm d}{\rm Lips} 
\\
&=&
{N\over{\cal O}_o}\int\ln\left({1+\omega_r \alpha_r+\omega_i \alpha_i+\omega_2 (\alpha_r^2+\alpha_i^2)\over 1+\Omega_2 (\alpha_r^2+\alpha_i^2)}\right)O_o
\ {\rm d}{\rm Lips} 
+ {\rm cst}
\end{eqnarray}
where the constant "${\rm cst}$" is irrelevant and where one introduced the notation:
\begin{eqnarray}
\omega_r={O_r\over O_o} \ \ \ ; \ \ \
\omega_i={O_i\over O_o} \ \ \ ; \ \ \
\omega_2={O_2\over O_o} \ \ \ ; \ \ \
\Omega_2={{\cal O}_2\over{\cal O}_o}
\end{eqnarray}
The inverse of the covariance matrix is given by:
\begin{equation}
\Lambda_{ij}=-\ {\partial^2 L(\alpha)\over\partial\alpha_i\partial\alpha_j}_{\mid \alpha_0}
\end{equation}
Evaluated for $\alpha_r^0=\alpha_i^0=0$, the $\Lambda_{rr}$ term boils down to:
\begin{eqnarray}
\Lambda_{rr}&=&
N\left(\int \omega_r^2 {O_o\over{\cal O}_0}\ {\rm d}{\rm Lips} -{2\over{\cal O}_o}\int O_2
\ {\rm d}{\rm Lips} 
+2\Omega_2\right)
=
N\langle \omega_r^2\rangle_{_0}
\end{eqnarray}
and similarly for the other terms.
One thus recovers the result of Eq.(\ref{MasterFormula}), with $\Omega_{r/i}={\cal O}_{r/i}/{\cal O}_0=0$.
In particular only $\omega_i$ and $\omega_r$ enter in the expression of $\Lambda_{ij}$.

From the $\omega_r$ and $\omega_i$ distributions, as provided from the narrow width Monte Carlo simulation, one obtains:
\begin{equation}
\sigma[\alpha_r]=\sqrt{\Lambda_{11}^{-1}}\simeq 0.01\ \ \ ; \ \ \ \sigma[\alpha_i]=\sqrt{\Lambda_{22}^{-1}}\simeq 0.01
\end{equation}
with 
a limited improvement due to beam polarizations.
On the other hand, we found that using the full matrix element, one improves the accuracy on the CP violating parameter 
by almost one order of magnitude with respect to~\cite{Ladinsky:1992vv}, which illustrates the potential gain the method can provide.

 \end{document}